\documentclass[aps,preprint,showpacs]{revtex4}
\usepackage{amssymb}
\usepackage{bm}
\usepackage{amsmath}
\usepackage{graphicx}
\usepackage{subfigure}

\begin{document}

\title{Gap polariton solitons}
\author{A. V. Gorbach$^{1}$, B. A. Malomed$^{2}$, D. V. Skryabin$^{1}$}
\affiliation{$^{1}$ Centre for Photonics and Photonic Materials, Department of Physics,
University of Bath, Bath BA2 7AY, UK}
\affiliation{$^{2}$ Department of Physical Electronics, Faculty of Engineering, Tel
Aviv University, Tel Aviv 69978, Israel}

\begin{abstract}
We report the existence, and study mobility and interactions of gap
polariton solitons in a microcavity with a periodic potential, where the
light field is strongly coupled to excitons. Gap solitons are formed due to
the interplay between the repulsive exciton-exciton interaction and cavity
dispersion. The analysis is carried out in an analytical form, using the
coupled-mode (CM) approximation, and also by means of numerical methods.
\end{abstract}

\maketitle

\section{Introduction}

The strong light-matter coupling in semiconductor microcavities has recently
attracted much attention \cite{KBM+2007}. In particular, the strong and fast
nonlinear response of microcavity exciton-polaritons has allowed to predict
and observe several important nonlinear effects, such as bistability \cite%
{BKE+2004,gip,CC2004}, parametric wave mixing \cite{gip,CC2004,SBS+2000},
superfluidity \cite{CC2004,bogol} and formation of solitons \cite%
{agr,YEL+2008,ESY+2009}. While the exciton-polariton nonlinearity is
defocusing due to the electrostatic repulsion of excitons, the effective
dispersion of the electromagnetic wave may be controlled in microcavities
with periodic potentials. The latter can be created by mirror patterning
\cite{LKU+2007}, or by way of surface acoustic waves \cite{CON+2005,berlin}.
In either case, the periodic modulation of system parameters can be achieved
on the micron scale, leading to the emergence of gaps in the polariton
spectrum. Localized nonlinear modes with the Fourier transform residing
within the forbidden gaps of linear spectra are called \emph{gap solitons}
(GSs), also known as Bragg solitons. GSs may exist with any sign of the
nonlinearity. Photonic GSs have been extensively studied in fiber gratings
and planar optical lattices \cite{kivshar}. Matter-wave GSs have been
observed in the atomic condensate of $^{87}$Rb \cite{Markus}. From the vast
literature on solitons in periodic structures it is relevant here to mention
works which either considered cavity effects or where material excitations
played a crucial role. These include the soliton transmission through
resonantly absorbing Bragg gratings \cite{mant,malom,maim}, control of
electro-magnetically induced transparency using photonic bandgaps \cite%
{lukin}, and light-only solitons in microcavities with photonic crystals
\cite{skryab,egorov}. The aim of this work is to initiate studies of the
exciton-polariton GSs, which are half-light half-matter nonlinear
excitations, whose self-localization is supported by a periodic potential acting
on the photonic component.

\section{The polariton model and its linear properties}

Below we focus on the microcavity model, which disregards cavity losses,
aiming at the proof-of-the-principle demonstration of the existence and
robustness of gap polariton solitons in this setting. Effects of the
dissipation and introduction of a compensating gain may be important to the
experimental realization, and will be considered elsewhere. In the scaled
form, the equations for local amplitudes of the photon ($E$) and exciton ($%
\Psi $) fields are \cite{KBM+2007,agr}
\begin{eqnarray}
&\partial _{t}E&-i(\partial _{x}^{2}+\partial _{y}^{2})E-iU(x,y)E=i\Psi \;,
\label{eqE} \\
&\partial _{t}\Psi &+i|\Psi |^{2}\Psi =iE\;.  \label{eqPsi}
\end{eqnarray}%
In these equations, the time and coordinates $x,y$ are measured,
respectively, in units of $1/\Omega _{R}$ and $1/k\sqrt{\omega /(2\Omega
_{R})}$, where $\Omega _{R}$ is the Rabi frequency, $\omega $ and $%
k=cn/\omega $ are the pump frequency and wavenumber, with $n$ being the
refractive index. Further, $\Omega _{R}|E|^{2}/g$ and $\Omega _{R}|\Psi
|^{2}/g$ are numbers of photons and excitons per unit area \cite{YEL+2008},
and $g$ is the exciton-exciton interaction constant. Taking typical
parameters of a microcavity based on a single InGaAs/GaAs quantum well, $%
\hbar \Omega _{R}\simeq 2.5\;$meV, $\hbar g\simeq 10^{-4}\;$eV$\cdot \mathrm{%
\mu }$m$^{2}$ \cite{BKE+2004,CC2004}, one finds that $|E|^{2}=1$ corresponds
to the electromagnetic field with intensity $\simeq 10\;$kW/cm$^{2}$, while
the time and length units translate into $\simeq 0.6\;$ps and $\simeq 1\;%
\mathrm{\mu }$m, respectively \cite{YEL+2008, ESY+2009}.

The form of Eqs. (\ref{eqE}) and (\ref{eqPsi}), with zero detuning between
them, assumes identical resonance frequencies of photons and excitons. Note
that the separation between the nonlinearity and diffraction in these
equations resembles the phenomenological model introduced earlier in Ref.
\cite{Zafrany}.

The last term on the left-hand side of Eq.~(\ref{eqE}) is the lattice
potential induced by the periodic modulation of the cavity resonance \cite%
{LKU+2007}. First, we consider one-dimensional (1D) case, with $y$%
-independent fields, and take potential
\begin{equation}
U(x)=\epsilon \cos (2k_{0}x),  \label{pot_1d}
\end{equation}%
where $\epsilon $ is the depth of the potential and $\pi /k_{0}$ is its period,
so that the first Brillouin zone for polariton momentum $k$ is $-k_{0}\leq
k\leq k_{0}$. A 2D model, in which $U(x,y)$ is
periodic in $x$ and localized in the $y$ direction, is considered
towards the end of the paper.

\begin{figure}[tbp]
\centering
\includegraphics[width=0.21\textwidth]{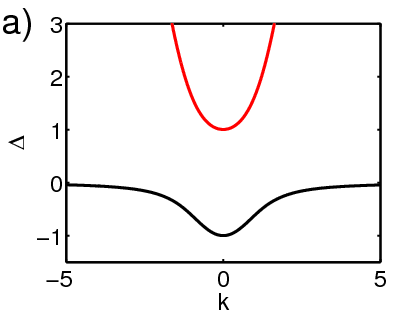} %
\includegraphics[width=0.21\textwidth]{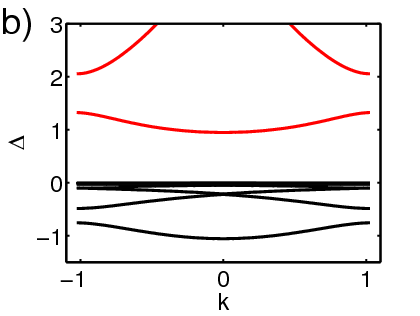} %
\includegraphics[width=0.21\textwidth]{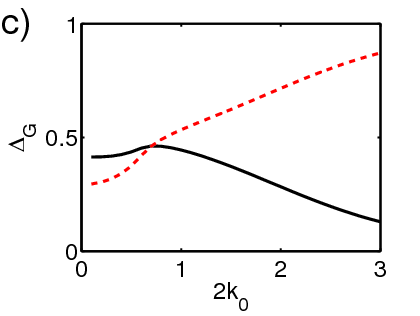} %
\includegraphics[width=0.21%
\textwidth]{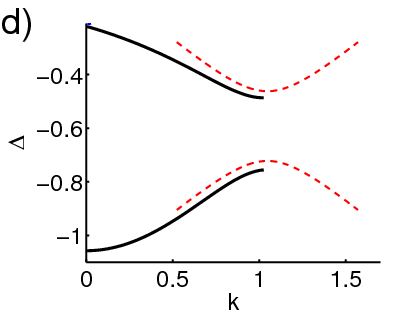}
\caption{(Color online) (a) The polariton spectrum of the homogeneous
microcavity ($\protect\epsilon =0$). Black (red) line corresponds to the
lower (upper) polariton branch. (b) The spectrum of the microcavity equipped
with the periodic potential: $\protect\epsilon =1$, $k_{0}=\protect\pi /3$.
(c) $k_{0}$-dependence of the width of the principal gaps in the lower
(black solid lines) and upper (dashed red lines) polariton branches, for $%
\protect\epsilon =1$. (d) The polariton spectrum in a vicinity of the
primary gap of the lower polariton branch. Solid black lines show the the
exact spectrum, and dashed red lines represent the approximation obtained
from Eqs.~(\protect\ref{coupled-mode1}), (\protect\ref{coupled-mode2}).}
\label{figbands}
\end{figure}

Without the lattice potential, $\epsilon =0$, solutions to the linearized
version of Eqs. (\ref{eqE}) and (\ref{eqPsi}) are sought for as $E,\Psi \sim
e^{ikx-i\Delta t}$, which yields the spectrum consisting of two branches,%
\begin{equation}
\Delta _{\pm }(k)=\frac{k^{2}}{2}\pm \sqrt{\frac{k^{4}}{4}+1}\;,
\label{spect_hom}
\end{equation}%
see Fig.~\ref{figbands}(a). The addition of the lattice potential splits
this spectrum into multiple bands with the zone folding happening at $%
k=k_{0} $, see Fig.~\ref{figbands}(b). Gaps between the bands are getting
wider for deeper lattices (larger $\epsilon $). Unusually, the choice of $%
k_{0}$ also affects the gap widths, see Fig.~\ref{figbands}(c). This
happens because the curvature of the exciton-polariton dispersion
without the periodic potential, see Eq. (\ref{spect_hom}), strongly
depends on $k$. In contrast, the photonic dispersion is parabolic,
and it is modified by the periodic potential in such a way that the
width of the primary gap is $\Delta _{G}=\epsilon $, being obviously
independent of $k_{0}$. In Fig.~\ref{figbands}(c) we plot the widths
of the primary gaps in the two dispersion
branches ($\Delta _{\pm }$) as functions of $k_{0}$. For relatively small $%
k_{0}$, the gaps in the upper and lower polariton branches are approximately
the same. Increasing $k_{0}$, the dispersion of the upper polariton branch
tends to its photonic (light-only) limit, hence the width of the primary gap
in the upper branch increases and tends to $\Delta _{G}=\epsilon $. For the
lower polariton branch, which is the practically important one (see below), $%
\Delta _{G}$ first increases and then drops to zero for large $k_{0}$,
see Fig.~\ref{figbands}(c). Below we focus on GSs residing in the principal
gap on the lower polariton branch. Our choice of $k_{0}$ throughout this
paper is $\pi /3$, which corresponds to modulation period $\simeq 3\mu $m
and matches the experimental conditions of Ref. \cite{LKU+2007}.

\section{The existence and robustness of gap solitons in the 1D geometry}

In order to address the existence of GSs in the model, we first apply the
coupled-mode (CM)\ approximation, which is known to lead to explicit
analytical results \cite{kivshar}, which are then compared to full numerical
solutions of Eqs. (\ref{eqE}), (\ref{eqPsi}). To this end, we introduce $%
\overrightarrow{F}=[E,\Psi ]^{T}$ and assume
\begin{eqnarray}
\overrightarrow{F} &=&\vec{\alpha _{0}}\left[ C_{+}(x,t)\exp (-i\Delta
_{0}t+ik_{0}x)+\right.   \notag \\
&&\left. C_{-}(x,t)\exp (-i\Delta _{0}t-ik_{0}x)\right] \;,
\label{weak_lattice_ansatz}
\end{eqnarray}%
where $\Delta _{0}=\Delta _{-}(k_{0})$ is the frequency as given by Eq. (\ref%
{spect_hom}) in the middle of the gap, $\vec{\alpha _{0}}=[-\Delta
_{0},1]^{T} $ is the polariton eigenvector in the absence of the
nonlinearity and periodic potential, and $C_{+}$ and $C_{-}$ are
slowly varying functions.
The substitution of Eq. (\ref{weak_lattice_ansatz}) into Eqs.~(\ref{eqE}), (%
\ref{eqPsi}) and manipulations similar to those known in the context of the
CM equations for fiber Bragg gratings \cite{kivshar,yulin}, we derive the CM
equations corresponding to the present setting:
\begin{gather}
i\left( \Delta _{0}^{2}+1\right) \partial _{t}C_{+}+2ik_{0}\Delta
_{0}^{2}\partial _{x}C_{+}+\kappa C_{-}  \notag \\
-\left( \left\vert C_{+}\right\vert ^{2}+2\left\vert C_{-}\right\vert
^{2}\right) C_{+}=0,  \label{coupled-mode1} \\
i\left( \Delta _{0}^{2}+1\right) \partial _{t}C_{-}-2ik_{0}\Delta
_{0}^{2}\partial _{x}C_{-}+\kappa C_{+}  \notag \\
-\left( \left\vert C_{-}\right\vert ^{2}+2\left\vert C_{+}\right\vert
^{2}\right) C_{-}=0,  \label{coupled-mode2}
\end{gather}%
where $\kappa =\epsilon \Delta _{0}^{2}/2$ is the coefficient of the
Bragg-reflection-induced linear coupling between the counterpropagating
waves. Figure \ref{figbands}(d) compares the spectra found from the full
model and from the CM equations. The good agreement between the two persists
for $\epsilon /k_{0}^{2}\lesssim 1$, i.e., for relatively weak potentials.

Using variables $\xi \equiv x/\left( 2k_{0}\Delta _{0}^{2}\right) $ and $%
\tau \equiv t/(1+\Delta _{0}^{2})$, we obtain from Eqs.
(\ref{coupled-mode1}) and (\ref{coupled-mode2}) explicit solutions
for GSs \cite{kivshar},
\begin{equation}
C_{\pm }=u_{\pm }(\eta )\exp \left[ iq(\eta )\pm is(\eta )-i\delta \sqrt{%
1-V^{2}}\tau \right] ,  \label{sol_ansatz_weak_lattice2}
\end{equation}%
\begin{eqnarray}
u_{+}^{2} &=&\frac{2(\kappa ^{2}-\delta ^{2})(1+V)\sqrt{1-V^{2}}}{\left(
3-V^{2}\right) \kappa \cosh (2\sqrt{\kappa ^{2}-\delta ^{2}}\eta )-\delta }~,
\label{soliton_u} \\
u_{-}^{2} &=&\frac{1-V}{1+V}u_{+}^{2},~q=V\delta \eta +\frac{4V}{3-V^{2}}s~,
\label{soliton_q} \\
s &=&-\tan ^{-1}\left[ \sqrt{\frac{\kappa +\delta }{\kappa -\delta }}\tanh
\left( \sqrt{\kappa ^{2}-\delta ^{2}}\eta \right) \right] ,
\label{soliton_s}
\end{eqnarray}%
where $\delta $ is the frequency detuning relative to the gap center, $|V|<1$
is the soliton velocity, and $\eta \equiv (\xi -V\tau )/\sqrt{1-V^{2}}$.

\begin{figure}[tbp]
\centering
\includegraphics[width=0.22%
\textwidth]{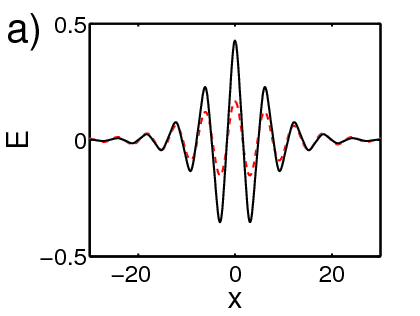} %
\includegraphics[width=0.22%
\textwidth]{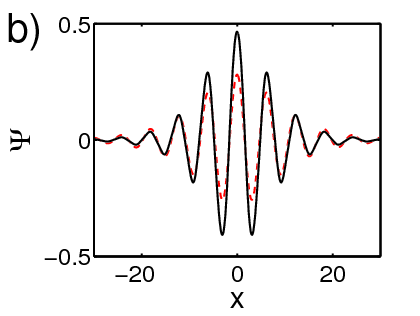} %
\caption{(Color online) Numerical (solid black lines) and analytical (dashed
red lines) solutions for gap solitons at $\protect\epsilon =1$, $k_{0}=%
\protect\pi /3$, $\Delta =-0.7$. }
\label{figsolweakcompare}
\end{figure}

In Fig. \ref{figsolweakcompare} we compare the analytical stationary
solitons given by Eqs. (\ref{sol_ansatz_weak_lattice2}) and their
counterparts found numerically from Eqs. (\ref{eqE}) and (\ref{eqPsi}). The
overall agreement is reasonable. The main source of the error is the
inaccuracy of ratio $|E|/|\Psi |$, which, in the framework of the CM
approximation, is taken as per the linear eigenvector at $\epsilon =0$ and $%
\Delta =\Delta _{0}$, therefore it is assumed to remain constant while the
soliton's spectrum is shifting within the gap, following a variation of $%
\delta $. However, numerical solutions show a tangible dependence of the $%
|E|/|\Psi |$ ratio on the soliton frequency $\delta $.

Next, we check the stability of the numerically found GS solutions. To this
end, we perturb them by setting $\left[ E,\Psi \right] ^{T}=\left[
E_{0}(x)+e(x,t),\Psi _{0}(x)+\psi (x,t)\right] ^{T}\cdot \exp (-i\Delta
_{0}t)$, and linearize Eqs.~(\ref{eqE}), (\ref{eqPsi}) assuming that
perturbations $e\left( x,t\right) $ and $\psi \left( x,t\right) $ are small.
This yields
\begin{eqnarray}
&&\partial _{t}\vec{y}=\hat{L}\vec{y},~\vec{y}\equiv \left[ e,e^{\ast },\psi
,\psi ^{\ast }\right] ^{T},  \label{eqPerturb} \\
\hat{L}= &&i\left[
\begin{array}{cccc}
\label{matrixL}L_{e} & 0 & 1 & 0 \\
0 & -L_{e} & 0 & -1 \\
1 & 0 & \Delta _{0}-2|\Psi _{0}|^{2} & -\Psi _{0}^{2} \\
0 & -1 & (\Psi _{0}^{2})^{\ast } & -\Delta _{0}+2|\Psi _{0}|^{2}%
\end{array}%
\right] ,  \notag
\end{eqnarray}%
where $\hat{L}_{e}\equiv \Delta _{0}+\partial _{x}^{2}+\epsilon \cos
(2k_{0}x)$. Then, we seek solutions to Eq. (\ref{eqPerturb}) as $\vec{y}%
(x,t)=\exp (\lambda t)\vec{y}_{0}(x)$. The existence of eigenvalues with $%
\mathrm{Re}\left\{ \lambda \right\} >0$ implies instability.

\begin{figure}[tbp]
\centering
\includegraphics[width=0.22\textwidth]{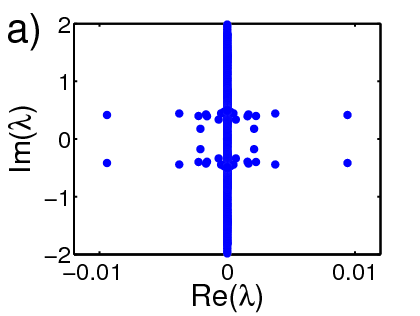} %
\includegraphics[width=0.22%
\textwidth]{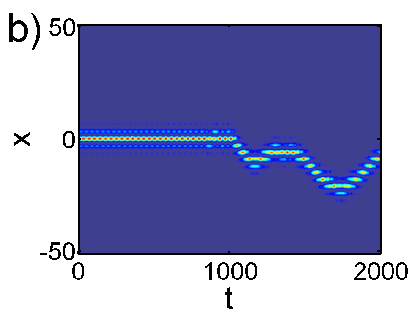}
\caption{(a) The stability spectrum for the soliton with $\Delta =-0.5$ and
other parameters as in Fig.~\protect\ref{figsolweakcompare}. Eigenvalues
with $\mathrm{Re}\left\{ \protect\lambda \right\} >0$ correspond to the
instability. (b) Evolution of an unstable soliton perturbed by random noise.}
\label{figstability}
\end{figure}

We have found that results of the stability analysis for the full polariton
model qualitatively coincide with the known stability properties of the GSs
obeying CM equations (\ref{coupled-mode1}), (\ref{coupled-mode2}) \cite%
{stability,yulin}. In particular, the gap polariton solitons are stable in
the lower half of the gap, and feature various instabilities in the upper
half, see Fig.~\ref{figstability}(a). Figure~\ref{figstability}(b)
demonstrates that the instability (if any) initiated by random perturbations
causes the soliton to ramble erratically across the lattice.

\begin{figure}[tbp]
\centering
\includegraphics[width=0.22\textwidth]{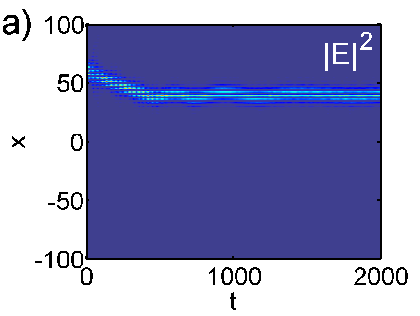} %
\includegraphics[width=0.22\textwidth]{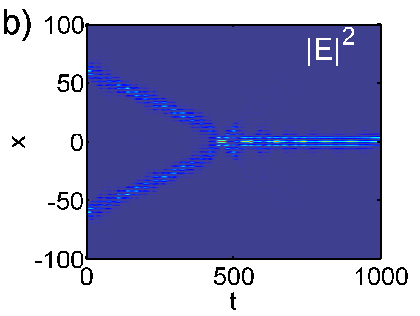} %
\includegraphics[width=0.22\textwidth]{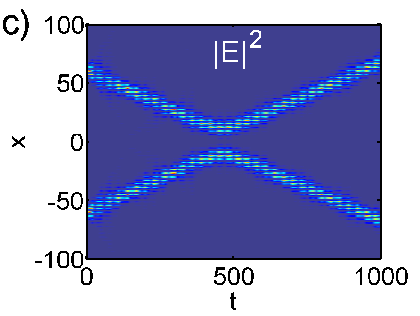}
\includegraphics[width=0.22\textwidth]{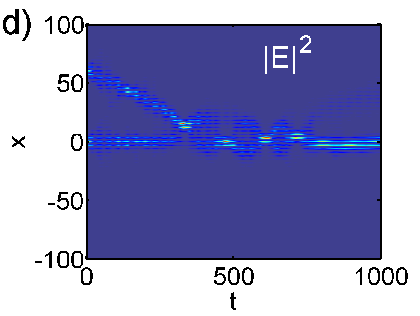}
\caption{ (a) Pinning of the soliton, which was initially moving with a
relatively small velocity, $V=0.1$ (b) Merger of colliding in-phase solitons
moving with velocities $V=\pm 0.2$. (c) Rebound of $\protect\pi $%
-out-of-phase solitons colliding with the same velocities as in (b). (d)
Collision between the soliton moving at $V=0.2$ and a quiescent one.
Parameters of the potential are the same as in Fig.~1(b),(d), and $\Delta
=-0.6$. }
\label{figsolmove}
\end{figure}

We have also checked numerically mobility and collisions of the GSs in the
full model. Figure~\ref{figsolmove}(a) shows a soliton which initially moves
through the lattice as prescribed by the CM approximation, but eventually
gets pinned around one of the lattice sites. Outcomes of collisions between
the solitons are sensitive to both initial velocities and the relative phase
of the interacting solitons, as shown in Figs. \ref{figsolmove}(b)-(d).
Collisions between solitons with opposite velocities result in merger of
in-phase soliton pairs, and rebound of the solitons with the phase different
of $\pi $, see Figs. \ref{figsolmove}(b)-(c). Collisions of solitons with
different velocities can produce a plethora of outcomes, with one example
shown in Fig. \ref{figsolmove}(c).

\section{Gap solitons in the 2D geometry}

\begin{figure}[tbp]
\includegraphics[width=0.22\textwidth]{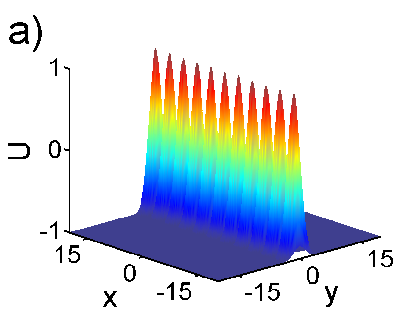} %
\includegraphics[width=0.22\textwidth]{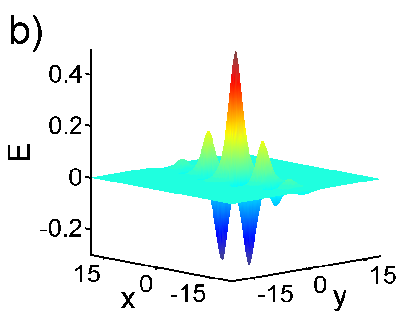} %
\caption{(a) The surface plot of the potential (\protect\ref%
{pot_quasi1d}) used in the 2D model, for $\protect\epsilon =1$, $k_{0}=0.85$%
, $w=2$ (the usual notation of the quantum theory requires to replace $U$ in
Eq.~(\ref{eqE}) by $-U$, so that the potential structure in (a) corresponds to
the periodic sequence of potential wells). (b)
The soliton's profile for $\Delta =-0.65$.}
\label{fig_quasi1D}
\end{figure}

To examine the relevance of the 1D model elaborated above to the 2D geometry
of practical interest, we consider a configuration where the periodic
potential acting in the $x$ direction is combined with a localized potential
applied along the $y$ coordinate:
\begin{equation}
U(x,y)=-\epsilon \left\{ 1-\exp \left[ -(y/w)^{2}\right] \left[ 1+\cos
(2k_{0}x)\right] \right\} \;.  \label{pot_quasi1d}
\end{equation}%
The shape of the above potential is shown in Fig. ~\ref{fig_quasi1D}(a). We
have found the corresponding 2D soliton solutions numerically, see Fig.~\ref%
{fig_quasi1D}(b), using a time-independent iteration method.

To check if the dynamics seen in the 1D case is retained in the 2D
configuration we have carried out a series of numerical experiments in
soliton collisions. The initial conditions were set as
\begin{equation}
E_{0}=\Psi _{0}=Ae^{-(x-x_{0})^{2}/w_{x}^{2}-y^{2}/w_{y}^{2}}\cos
(k_{0}x)e^{ikx}\;,  \label{ini_cond_channel}
\end{equation}%
where $A$, $w_{x}$ and $w_{y}$ have been chosen to approximate the
stationary profile of the numerically found solitons and $k$ is the initial
soliton momentum. The results of these simulations are shown in Figs.~\ref%
{fig_channel_collide_sin} and \ref{fig_channel_collide_anti}. Similar to the
1D case, the in-phase solitons tend to merge in the course of the collision,
while the out-of-phase solitons bounce back. The outcome of the collision of
the out-of-phase 2D solitons is similar to what was observed in the 1D
model. However, for the in-phase solitons the large-amplitude pattern
generated by the merger is, most often, unstable, splitting into two
quasi-solitons moving in opposite directions, see Fig.~\ref%
{fig_channel_collide_sin}. This pair is asymmetric, one of the emerging
quasi-solitons being usually larger and slower than the other.

\begin{figure}[tbp]
\includegraphics[width=0.45\textwidth]{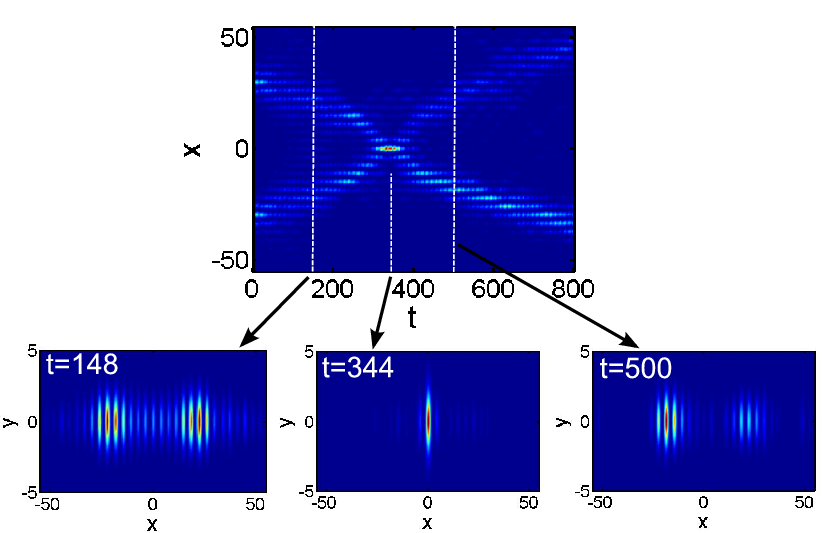}
\caption{Collision of in-phase solitons in the 2D geometry. All panels show $%
|E(x,y,t|^{2}$. The top panel shows dynamics in the $(x,t)$-plane along $y=0$%
. Bottom panels show the light intensity in the $(x,y)$-plane for fixed
time. Parameters are the same as in Fig.~\protect\ref{fig_quasi1D}. Initial
conditions for each soliton are chosen as in Eq.~(\protect\ref%
{ini_cond_channel}), with $A=0.5$, $w_{x}=5$, $w_{y}=3$, $k=\pm k_{0}/20$.}
\label{fig_channel_collide_sin}
\end{figure}

\begin{figure}[tbp]
\includegraphics[width=0.45\textwidth]{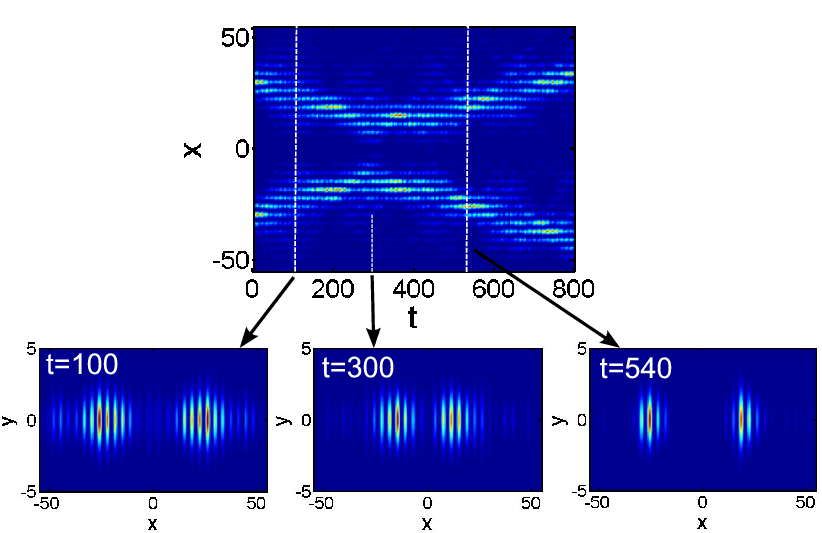}
\caption{The same as in Fig.~\protect\ref{fig_channel_collide_sin}, but for
the $\protect\pi $ out-of-phase solitons.}
\label{fig_channel_collide_anti}
\end{figure}

\section{Summary}

We have predicted the existence and studied stability, mobility and
interactions of gap polariton solitons in microcavities equipped with
periodic photonic potentials. The 1D model has been studied using the CM
(coupled-mode) approach, which yields analytical solutions for the solitons,
and also by way of the numerical solution of the full system of equations
for the photonic and excitonic components of polaritons. Furthermore, we
have studied the two-dimensional microcavity with the periodic potential
along one dimension and the trapping potential along the other.

B.A.M. appreciates hospitality of the Department of Physics at the
University of Bath (UK). A.V.G. and D.V.S. acknowledge support from EPSRC 
(grant EP/D079225/1).


\end{document}